\documentclass[a4paper,11pt]{article}

\usepackage{jheppub_mod} 
\usepackage[T1]{fontenc} 

\usepackage{amsmath,booktabs}
\usepackage{graphicx}
\usepackage{xspace}
\usepackage{color}
\pagecolor{white}
\usepackage[dvipsnames]{xcolor}
\usepackage{url}
\usepackage[utf8]{inputenc}
\usepackage{epstopdf}
\usepackage{hyperref}
\usepackage{multirow}
\usepackage{gensymb}

\usepackage{ifmtarg}

\newcommand{\beq}{\begin{equation}}
\newcommand{\eeq}{\end{equation}}

\newcommand{\M}{\mathcal{M}}

\newcommand{\GeV}{\,\text{GeV}}

\renewcommand{\Im}{\text{Im}\,}

\newcommand{\F}{\mathcal{F}}

\newcommand{\mP}{M_P}
\newcommand{\mS}{M_S}
\newcommand{\mA}{M_A}
\newcommand{\mT}{M_T}
\newcommand{\Imts}{\text{Im}_t^s\,}
\newcommand{\Imtu}{\text{Im}_t^u\,}
\newcommand{\Imst}{\text{Im}_s^t\,}
\newcommand{\Imsu}{\text{Im}_s^u\,}
\newcommand{\Imus}{\text{Im}_u^s\,}
\newcommand{\Imut}{\text{Im}_u^t\,}

\makeatletter
\newcommand{\Cr}[2]{\@ifmtarg{#2}{\mathcal{C}_{#1}}{\mathcal{C}_{#1}\big[#2\big]}}
\makeatother

\allowdisplaybreaks[1]

\title{An optimized basis for hadronic light-by-light scattering}

\author[a]{Martin Hoferichter,}
\author[b,c]{Peter Stoffer,}
\author[a]{and Maximilian Zillinger}

\affiliation[a]{Albert Einstein Center for Fundamental Physics, Institute for Theoretical Physics, University of Bern, Sidlerstrasse 5, 3012 Bern, Switzerland}

\affiliation[b]{Physik-Institut, Universit\"at Z\"urich, Winterthurerstrasse 190, 8057 Z\"urich, Switzerland}

\affiliation[c]{Paul Scherrer Institut, 5232 Villigen PSI, Switzerland}

\preprint{PSI-PR-24-08, ZU-TH 11/24}

\emailAdd{hoferichter@itp.unibe.ch}
\emailAdd{stoffer@physik.uzh.ch}
\emailAdd{zillinger@itp.unibe.ch}

\abstract{We present a new basis for the hadronic light-by-light (HLbL) tensor that is optimized for the evaluation of narrow-resonance contributions to HLbL scattering in the anomalous magnetic moment of the muon. As main advantage, kinematic singularities are manifestly absent for pseudoscalar, scalar, and axial-vector states, while the remaining singularities for tensor resonances are minimized, even avoided for special cases, and simple crossing relations among the scalar functions maintained. We scrutinize the properties of this new basis for the scalar-QED pion box, demonstrating that the partial-wave convergence even slightly improves compared to our previous work, and discuss the physical sum rules that ensure basis independence of the HLbL contribution. Finally, we provide explicit expressions for narrow (pseudo-)scalar, axial-vector, and tensor intermediate states in terms of their respective transition form factors.}

\begin{document}
\maketitle
	
\section{Introduction}
\label{sec:intro}

The current experimental value for the anomalous magnetic moment of the muon $a_\mu=(g-2)_\mu/2$, is dominated by the Fermilab experiment~\cite{Muong-2:2023cdq,Muong-2:2024hpx,Muong-2:2021ojo,Muong-2:2021ovs,Muong-2:2021xzz,Muong-2:2021vma,Muong-2:2006rrc}
\beq
\label{eq:amuExp}
	a_\mu^\text{exp}=116\,592\,059(22)\times 10^{-11},
\eeq
having reached an impressive precision of $0.19\,\text{ppm}$. While the need for theory improvement is most evident for hadronic vacuum polarization,\footnote{The Standard-Model (SM) prediction from Refs.~\cite{Aoyama:2020ynm,Aoyama:2012wk,Aoyama:2019ryr,Czarnecki:2002nt,Gnendiger:2013pva,Davier:2017zfy,Keshavarzi:2018mgv,Colangelo:2018mtw,Hoferichter:2019gzf,Davier:2019can,Keshavarzi:2019abf,Hoid:2020xjs,Kurz:2014wya,Melnikov:2003xd,Colangelo:2014dfa,Colangelo:2014pva,Colangelo:2015ama,Masjuan:2017tvw,Colangelo:2017qdm,Colangelo:2017fiz,Hoferichter:2018dmo,Hoferichter:2018kwz,Gerardin:2019vio,Bijnens:2019ghy,Colangelo:2019lpu,Colangelo:2019uex,Blum:2019ugy,Colangelo:2014qya} suggests a $5.1\sigma$ difference, but recent tensions both within data-driven evaluations~\cite{Davier:2017zfy,Keshavarzi:2018mgv,Colangelo:2018mtw,Hoferichter:2019gzf,Davier:2019can,Keshavarzi:2019abf,Hoid:2020xjs,Stamen:2022uqh,Colangelo:2022vok,Colangelo:2022prz,Hoferichter:2023sli,Hoferichter:2023bjm,Davier:2023fpl} (most notably in the context of the CMD-3 measurement of $e^+e^-\to\pi^+\pi^-$~\cite{CMD-3:2023alj,CMD-3:2023rfe} and potentially related to radiative corrections~\cite{Campanario:2019mjh,Ignatov:2022iou,Colangelo:2022lzg,Monnard:2021pvm,Abbiendi:2022liz,BaBar:2023xiy}) and with lattice QCD~\cite{Borsanyi:2020mff,Ce:2022kxy,ExtendedTwistedMass:2022jpw,FermilabLatticeHPQCD:2023jof,RBC:2023pvn} will have to get resolved to enable a meaningful comparison with experiment.} the uncertainty of the hadronic-light-by-light (HLbL) contribution, $a_\mu^\text{HLbL} = 92(19) \times 10^{-11}$~\cite{Aoyama:2020ynm,Melnikov:2003xd,Masjuan:2017tvw,Colangelo:2017qdm,Colangelo:2017fiz,Hoferichter:2018dmo,Hoferichter:2018kwz,Gerardin:2019vio,Bijnens:2019ghy,Colangelo:2019lpu,Colangelo:2019uex,Pauk:2014rta,Danilkin:2016hnh,Jegerlehner:2017gek,Knecht:2018sci,Eichmann:2019bqf,Roig:2019reh}, is at the same level as the experimental uncertainty in Eq.~\eqref{eq:amuExp}, and thus needs to be improved further, by at least a factor $2$, to ensure that the reach of $a_\mu$ as a precision test of physics beyond the SM will not be limited by HLbL scattering~\cite{Muong-2:2015xgu,Colangelo:2022jxc}. Work in this direction is in progress both in lattice QCD~\cite{Chao:2021tvp,Chao:2022xzg,Blum:2023vlm} and with data-driven methods, the latter including the derivation of higher-order short-distance constraints~\cite{Bijnens:2020xnl,Bijnens:2021jqo,Bijnens:2022itw}, their implementation~\cite{Leutgeb:2019gbz,Cappiello:2019hwh,Knecht:2020xyr,Masjuan:2020jsf,Ludtke:2020moa,Colangelo:2021nkr,Leutgeb:2021mpu,Leutgeb:2022lqw,Colangelo:2024xfh}, and the application of dispersion relations.  
In this context, improvements of the $\eta$, $\eta'$ pole contributions~\cite{Holz:2015tcg,Holz:2022hwz,ExtendedTwistedMass:2022ofm,Gerardin:2023naa} will conclude the evaluation of light pseudoscalars, and $S$-wave two-meson intermediate states~\cite{Colangelo:2017qdm,Colangelo:2017fiz,Stamen:2022uqh}, in which light scalar resonances appear as unitarity corrections~\cite{Danilkin:2021icn,Deineka:2023nhu}, are already well under control. Accordingly, recent efforts have concentrated on subleading effects in the $(1\text{--}2)\GeV$ region, where axial-vector and tensor resonances constitute the dominant features in the open hadronic channels. In the dispersive formalism of Ref.~\cite{Colangelo:2017fiz}, the evaluation of such higher-spin contributions is affected by kinematic singularities, which are only guaranteed to cancel for the entire HLbL tensor, but not necessarily for each individual contribution. In Ref.~\cite{Colangelo:2021nkr} we already remarked that it is possible to construct a basis for HLbL scattering in which these singularities are manifestly absent for axial-vector intermediate states, so that their evaluation proceeds in a straightforward manner once input for the respective transition form factors (TFFs)~\cite{Rudenko:2017bel,Milstein:2019yvz,Hoferichter:2020lap,Zanke:2021wiq,Hoferichter:2023tgp} is available, and the remaining challenge concerns the interplay with short-distance constraints. The first part of this paper is devoted to the details of this construction. 

In particular, we studied systematically to what extent further optimizations for tensor resonances are possible. Assuming the same set of sum rules (SRs) as in Ref.~\cite{Colangelo:2017fiz}, we find that it is not possible to remove all kinematic singularities. However, in special cases in which only some of the TFFs are non-zero, observables not affected by kinematic singularities can be defined, e.g., when only $\F_1^T$ is present, as in the quark model of Ref.~\cite{Schuler:1997yw}. We imagine such benchmarks to be valuable in the context of the alternative dispersive formalism in triangle kinematics recently worked out in Ref.~\cite{Ludtke:2023hvz} and designed to allow for a general evaluation of higher-spin contributions, both in a narrow-resonance approximation and for general $\gamma^*\gamma^*\to\pi\pi$ $D$-waves~\cite{Garcia-Martin:2010kyn,Hoferichter:2011wk,Moussallam:2013una,Hoferichter:2013ama,Danilkin:2018qfn,Hoferichter:2019nlq,Danilkin:2019opj}.  
Given that phenomenological information on the tensor TFFs is limited, the combination of these complementary approaches should prove most advantageous for a robust evaluation of tensor contributions.  In Sec.~\ref{sec:HLbL_tensor} we present our optimized basis and discuss the details of its construction. 
 
Next, it is important to verify that the optimization for narrow-resonance contributions does not impede the partial-wave convergence of the pion-box contribution, which subsumes all terms with pion-pole left-hand cuts, as in this case the SRs are fulfilled exactly. This point is addressed in Sec.~\ref{sec:PW}, where we compare the convergence properties with our previous basis and demonstrate even a slight improvement, especially as regards saturation of the full result by the first few partial waves. In Sec.~\ref{sec:narrow_resonances} we give the explicit expressions for narrow-resonance contributions to the HLbL tensor, in terms of their TFFs, and all our main results are provided in a comprehensive \textsc{Mathematica} notebook. 
The main findings and future applications are summarized in Sec.~\ref{sec:summary}.

\section{HLbL tensor}
\label{sec:HLbL_tensor}

\subsection{BTT formalism and singly-on-shell basis}

Throughout this article, we follow the conventions of Ref.~\cite{Colangelo:2017fiz} for the HLbL tensor. That is, we start from a Bardeen--Tung--Tarrach (BTT)~\cite{Bardeen:1968ebo,Tarrach:1975tu} decomposition of the form
\beq
\label{decomposition}
\Pi^{\mu\nu\lambda\sigma} = \sum_{i=1}^{54} T_i^{\mu\nu\lambda\sigma} \Pi_i, 
\eeq
with Lorentz structures $T_i^{\mu\nu\lambda\sigma}$ as in Ref.~\cite{Colangelo:2017fiz} and scalar functions $\Pi_i$. The photon--photon scattering process is written in the conventions 
\beq
\gamma^*(q_1,\mu)\gamma^*(q_2,\nu)\to\gamma^*(-q_3,\lambda)\gamma^*(q_4,\sigma),
\eeq
with momenta $q_1+q_2+q_3=q_4$ and Lorentz indices as indicated.  The Mandelstam variables
\beq
s=(q_1+q_2)^2,\qquad t=(q_1+q_3)^2,\qquad u=(q_2+q_3)^2, 
\eeq
fulfill $s+t+u=\sum_i q_i^2$.
Dispersion relations are derived in these general kinematics, and only in the end is the limit $q_4\to 0$ taken, upon which $s=q_3^2$, $t=q_2^2$, and $u=q_1^2$. The crossing properties of various quantities are expressed in terms of the operators
\beq
\label{crossing}
\Cr{12}{f} = f(\mu \leftrightarrow \nu, q_1 \leftrightarrow q_2), 
\eeq
and likewise for the other photons. 

The HLbL master formula reads
\beq
	\label{eq:MasterFormulaPolarCoord}
	a_\mu^\text{HLbL} = \frac{\alpha^3}{432\pi^2} \int_0^\infty d\Sigma\, \Sigma^3 \int_0^1 dr\, r\sqrt{1-r^2} \int_0^{2\pi} d\phi \,\sum_{i=1}^{12} T_i(\Sigma,r,\phi) \bar\Pi_i(q_1^2,q_2^2,q_3^2) ,
\eeq
where the virtualities $Q_i^2=-q_i^2$ are parameterized via~\cite{Eichmann:2015nra}
\begin{align}
\label{Qi}
		Q_1^2 &= \frac{\Sigma}{3} \left( 1 - \frac{r}{2} \cos\phi - \frac{r}{2}\sqrt{3} \sin\phi \right), \notag\\
		Q_2^2 &= \frac{\Sigma}{3} \left( 1 - \frac{r}{2} \cos\phi + \frac{r}{2}\sqrt{3} \sin\phi \right),\notag \\
		Q_3^2 &= \frac{\Sigma}{3} \left( 1 + r \cos\phi \right).
\end{align}
The kernel functions $T_i(\Sigma,r,\phi)$ are listed in Ref.~\cite{Colangelo:2017fiz}, they are of purely kinematic nature. The dynamical content of the theory is represented by the scalar functions, to be concrete, in the master formula~\eqref{eq:MasterFormulaPolarCoord} one needs 
\begin{align}
		\label{eq:PibarFunctions}
		\bar \Pi_1 &= \hat\Pi_1, \quad &
		\bar \Pi_2 &= \hat\Pi_2, \quad &
		\bar \Pi_3 &= \hat\Pi_4, \quad &
		\bar \Pi_4 &= \hat\Pi_5, \quad &
		\bar \Pi_5 &= \hat\Pi_7, \quad &
		\bar \Pi_6 &= \hat\Pi_9, \notag\\
		\bar \Pi_7 &= \hat\Pi_{10}, \quad &
		\bar \Pi_8 &= \hat\Pi_{11}, \quad &
		\bar \Pi_9 &= \hat\Pi_{17}, \quad &
		\bar \Pi_{10} &= \hat\Pi_{39}, \quad &
		\bar \Pi_{11} &= \hat\Pi_{50}, \quad &
		\bar \Pi_{12} &= \hat\Pi_{54},
\end{align}
where the $\hat \Pi_i$ are linear combinations of the original $\Pi_i$ from the decomposition~\eqref{decomposition}, evaluated in the $g-2$ limit $q_4\to 0$. Via crossing symmetry, they follow from six representatives $\hat \Pi_i$, $i\in\{1,4,7,17,39,54\}$, e.g.,
\begin{align}
	\label{eq:CrossingRelationsPiHat}
	\hat \Pi_2 &= \Cr{23}{\hat \Pi_1}, \quad &
	\hat \Pi_5 &= \Cr{23}{\hat \Pi_4},  \quad &
	\hat \Pi_9 &= \Cr{12}{\Cr{13}{\hat \Pi_7}},\notag\\ 
 \hat \Pi_{10} &= \Cr{23}{\hat\Pi_7}, \quad & 
	\hat \Pi_{11} &= \Cr{13}{\hat \Pi_{17}}, \quad &  
	\hat \Pi_{50} &= -\Cr{23}{\hat \Pi_{54}}, 
\end{align}
and, in addition, one has the crossing relations
$\hat \Pi_{1,4,17} = \Cr{12}{\hat \Pi_{1,4,17}}$, $\hat \Pi_{54} = - \Cr{12}{\hat \Pi_{54}}$, while $\hat \Pi_{39}$ is symmetric under the exchange of either pair of virtualities. 

The main complications of deriving dispersion relations for the $\Pi_i$ ultimately arise from two properties of the BTT decomposition~\eqref{decomposition}. First, the $54$ elements do not form a basis, instead, one is forced to work with a redundant set of functions to avoid the appearance of kinematic singularities in the decomposition itself.  The resulting invariant functions $\Pi_i$ are suitable for a dispersive analysis, at the price of these Tarrach redundancies~\cite{Tarrach:1975tu}. Second, constraints on the spectral functions in the dispersion relations are most easily expressed in terms of helicity amplitudes, as these are the actual physical observables ($41$ in the fully off-shell case, $27$ for $q_4^2=0$), and the transition between helicity amplitudes and BTT scalar functions again introduces kinematic singularities. 
To address these issues, we constructed yet another set of functions, called $\check \Pi_i$, in Ref.~\cite{Colangelo:2017fiz}, which stand in a one-to-one correspondence to the $27$ singly-on-shell helicity amplitudes, fulfill unsubtracted fixed-$t$ dispersion relations, and in the $g-2$ limit reduce to the  $\hat\Pi_i$ necessary for Eq.~\eqref{eq:MasterFormulaPolarCoord}. Accordingly, this singly-on-shell basis  $\check \Pi_i$, which is free from Tarrach redundancies (and an extra ambiguity in four space-time dimensions~\cite{Eichmann:2014ooa} related to the Schouten identity~\cite{Ludtke:2023hvz}), in principle allows one to reconstruct the required scalar functions from helicity amplitudes accessible in experiment. 

Unfortunately, the kinematic singularities involved in the transition between singly-on-shell basis and helicity amplitudes do not cancel at every step of the calculation in a manifest way, only the entire HLbL tensor is guaranteed to be free of such kinematic singularities. On a technical level, the singularities cancel due to a set of SRs fulfilled by the $\check\Pi_i$, and these SRs may well connect different classes of contributions. In the basis put forward in Ref.~\cite{Colangelo:2017fiz}, they take the form of poles $1/\lambda_{123}$, $\lambda_{123}=\lambda(q_1^2,q_2^2,q_3^2)$, $\lambda(a,b,c)=a^2+b^2+c^2-2(ab+ac+bc)$, which, since $\lambda_{123}=-\Sigma^2(1-r^2)/3$, diverge along the border of the $g-2$ integration region. We demonstrated that for contributions that explicitly fulfill the SRs, such as the pion box, this singularity at $r=1$ can simply be subtracted, given that the residue is exactly zero on account of the SRs, and thereby were able to reproduce the full pion-box contribution from a partial-wave expansion. However, the disadvantage of this choice of basis was that single-particle intermediate states starting at angular momentum $J=1$ were affected by kinematic singularities and therefore not well-defined. The main purpose of this work is to use the freedom in the choice of basis to optimize the singularity structure and thereby push the appearance of kinematic singularities to higher $J$, to allow one to evaluate more narrow-resonance contributions in an unambiguous manner.

\subsection{Sum rules and narrow resonances}

There are $15$ physical SRs identified in Ref.~\cite{Colangelo:2017fiz}, of which, due to crossing symmetry, nine are independent. Their explicit form reads
\begin{align}
		\label{eq:SumRulesPicheck}
		\text{SR1 \ldots SR7}: \quad 0 &= \int ds' \Im \check\Pi_i(s') , \qquad i \in\{ 7, 8, 9, 16, 20,  22,  24\}, \notag\\
		\text{SR8}: \quad 0 &= \int ds' \Im\Big( \check\Pi_{11}(s') + \check\Pi_{18}(s') - \check\Pi_{19}(s') \Big) , \notag\\
		\text{SR9}: \quad 0 &= \int ds' \Im\Big( \check\Pi_{17}(s') - \check\Pi_{18}(s') + \check\Pi_{19}(s') \Big),
\end{align}
where in each case the notation is schematic and the full form of the SR integrals reads
\beq
0 = \frac{1}{\pi}\int_{s_0}^\infty ds' \Im \check\Pi_i\big(s',q_2^2,q_1^2+q_3^2-s'\big)-
\frac{1}{\pi}\int_{u_0}^\infty du' \Im \check\Pi_i\big(q_1^2+q_3^2-u',q_2^2,u'\big).
\eeq
$s_0$ and $u_0$ refer to the $s$- and $u$-channel thresholds, and the SRs are evaluated for fixed-$t$ kinematics at $t=q_2^2$. The $u$-channel discontinuities are most easily reconstructed from the $s$-channel ones via the crossing relations for the $\check\Pi_i$, which, for the functions appearing in the SRs, read
\begin{align}
\label{checkPi_crossing}
 \check \Pi_7&=\Cr{13}{\check\Pi_{12}}, &\quad    
 \check \Pi_8&=\Cr{13}{\check\Pi_{13}}, &\quad
 \check \Pi_9&=\Cr{13}{\check\Pi_{10}}, &\quad
 \check \Pi_{11}&=\Cr{13}{\check\Pi_{15}},\notag\\
 \check \Pi_{16}&=\Cr{13}{\check\Pi_{16}}, &\quad    
 \check \Pi_{17}&=-\Cr{13}{\check\Pi_{17}}, &\quad
 \check \Pi_{18}&=\Cr{13}{\check\Pi_{19}}, &\quad
 \check \Pi_{20}&=\Cr{13}{\check\Pi_{21}},\notag\\
  \check \Pi_{22}&=\Cr{13}{\check\Pi_{23}}, &\quad    
 \check \Pi_{24}&=\Cr{13}{\check\Pi_{24}}. & &&&
\end{align}
Accordingly, in the case $q_1^2=q_3^2$, the SRs for $\check\Pi_{16,24}$ are trivially fulfilled because the contributions from
the left- and the right-hand cuts cancel each other.

For narrow resonances, the imaginary parts $\Im\check\Pi_i$ collapse to $\delta$ functions $\delta(s-M^2)$, with resonance mass $M$, whose coefficient is determined by the respective TFFs, see Ref.~\cite{Hoferichter:2020lap} and Sec.~\ref{sec:narrow_resonances}. As an example, the SR for $\check\Pi_{24}$ (SR7) takes the form
\begin{align}
\label{SR7}
    0&=\sum_S\bigg(-\frac{q_1^2-q_2^2-q_3^2}{\mS^4}\F_2^S(q_1^2,q_2^2)\F_1^S(q_3^2,0)-\frac{q_1^2+q_2^2-q_3^2}{\mS^4}\F_2^S(q_2^2,q_3^2)\F_1^S(q_1^2,0)\bigg)\notag\\
    &+\sum_A\bigg(\frac{1}{\mA^4}\bigg[2\Big(2q_1^2\F_2^A(q_1^2,q_2^2)-(q_1^2+q_2^2+3q_3^2)\F_3^A(q_1^2,q_2^2)\Big)\F_1^A(q_3^2,0)\notag\\
    &\qquad +2\Big((-q_1^2+q_2^2+3q_3^2)\F_1^A(q_1^2,q_2^2)+q_3^2\F_a^A(q_1^2,q_2^2)\Big)\F_2^A(q_3^2,0)\bigg]\notag\\
    &\quad+\frac{1}{\mA^4}\bigg[2\Big(2q_3^2\F_3^A(q_2^2,q_3^2)-(3q_1^2+q_2^2+q_3^2)\F_2^A(q_2^2,q_3^2)\Big)\F_1^A(q_1^2,0)\notag\\
    &\qquad+2\Big((3q_1^2+q_2^2-q_3^2)\F_1^A(q_2^2,q_3^2)+q_1^2\F_a^A(q_2^2,q_3^2)\Big)\F_2^A(q_1^2,0)\bigg]+(J\geq 2).
\end{align}
Since pseudoscalars are not affected by Tarrach ambiguities, they do not contribute to the SRs, but all other intermediate states become intertwined. That is, in general such narrow-resonance contributions do not fulfill the SRs by themselves, and this is why the choice of the HLbL basis matters. In a prudently chosen basis in which the kinematic singularities are absent, some violation of the SRs can be tolerated to achieve a given accuracy, while in the presence of kinematic singularities the SRs have to be fulfilled exactly in order to ensure their cancellation.  

\subsection{Optimized basis}

To optimize the basis choice for narrow resonances, we proceeded as follows. First, we expressed all SRs, each of which corresponds to a degree of freedom in the optimization, in terms of the respective TFFs, as in Eq.~\eqref{SR7}, for scalars, axial-vectors, and tensors, where, in practice, it is easiest to go back to the original $15$ SRs and always keep pairs related by crossing symmetry (except for the SRs that map onto themselves upon crossing). Next, we considered linear combinations of the SRs that leave the scalar contributions unchanged, as these are already free from kinematic singularities in the basis of Ref.~\cite{Colangelo:2017fiz}. Moreover, we constructed linear combinations in which these remaining 13 SRs, when evaluated for narrow resonances, take a particularly simple form, e.g., removing all poles in $\lambda_{123}$ in favor of poles in $q_i^2$. Using this form of the SRs, we constructed a basis change for the $\check\Pi_i$ that optimizes the narrow resonances as follows: ten degrees of freedom are required to remove all singularities from the axial-vector contributions, leaving three degrees of freedom to simplify the expressions for tensor resonances. The linear combination leaving the scalar and axial-vector states invariant by demanding the coefficient of every combination of form factors $\F^{S}_{i}\F^{S}_{j}$, $\F^{A}_{i}\F^{A}_{j}$ to be identically zero reads
\begin{align}
0&=\big[f(q_{1}^{2},q_{2}^{2},q_{3}^{2})-8g(q_{1}^{2},q_{2}^{2},q_{3}^{2}) q_1^2 q_3^2 (q_{1}^2+q_{2}^{2}-q_{3}^2)\big]\times\text{SR1} \notag\\
&\quad+\big[f(q_{1}^{2},q_{2}^{2},q_{3}^{2})-16g(q_{1}^{2},q_{2}^{2},q_{3}^{2})q_{1}^{2}q_{2}^{2}q_{3}^{2}\big]\times\text{SR2} \notag\\
&\quad+\big[h(q_{1}^{2},q_{2}^{2},q_{3}^{2})\big]\times\text{SR3} \notag\\
&\quad+\big[f(q_{1}^{2},q_{2}^{2},q_{3}^{2})-16g(q_{1}^{2},q_{2}^{2},q_{3}^{2})q_{1}^{2}q_{2}^{2}q_{3}^{2}+h(q_{1}^{2},q_{2}^{2},q_{3}^{2})\big]\times\text{SR4} \notag\\
&\quad-\big[4g(q_{1}^{2},q_{2}^{2},q_{3}^{2})q_1^2(q_{1}^2-q_{2}^{2}-q_{3}^{2})\big]\times\text{SR5} \notag\\
&\quad+\big[g(q_{1}^{2},q_{2}^{2},q_{3}^{2})(3q_{1}^{2}-q_{2}^{2}+q_{3}^{2})\big]\times{\text{SR6}}\notag\\
&\quad+\big[g(q_{1}^{2},q_{2}^{2},q_{3}^{2}) \lambda_{123}\big]\times\text{SR7}\notag\\
&\quad-\big[4g(q_{1}^{2},q_{2}^{2},q_{3}^{2})q_1^2(\lambda_{123} + 2 q_3^2(q_1^2-q_2^2-q_3^2))\big]\times\text{SR8}\notag\\
&\quad+\big[f(q_{1}^{2},q_{2}^{2},q_{3}^{2})-h(q_{1}^{2},q_{2}^{2},q_{3}^{2})\big]\times\text{SR9}\notag\\
&\quad+\big[h(q_{1}^{2},q_{2}^{2},q_{3}^{2})+8g(q_{1}^{2},q_{2}^{2},q_{3}^{2})q_{1}^{2}q_3^2(q_{1}^{2}-q_{2}^{2}-q_{3}^2)\big]\times\Cr{13}{\text{SR1}}\notag\\
&\quad+\big[h(q_{1}^{2},q_{2}^{2},q_{3}^{2})-16g(q_{1}^{2},q_{2}^{2},q_{3}^{2})q_{1}^{2}q_{2}^{2}q_{3}^{2}\big]\times\Cr{13}{\text{SR2}}\notag\\
&\quad+\big[f(q_{1}^{2},q_{2}^{2},q_{3}^{2})\big]\times\Cr{13}{\text{SR3}}\notag\\
&\quad+\big[4g(q_{1}^{2},q_{2}^{2},q_{3}^{2})q_3^2(q_{1}^{2}+q_{2}^{2}-q_{3}^2)\big]\times\Cr{13}{\text{SR5}}\notag\\
&\quad+\big[g(q_{1}^{2},q_{2}^{2},q_{3}^{2})(q_{1}^{2}-q_{2}^{2}+3q_{3}^{2})\big]\times\Cr{13}{\text{SR6}}\notag\\
&\quad-\big[4g(q_{1}^{2},q_{2}^{2},q_{3}^{2})q_3^2(\lambda_{123} - 2 q_1^2(q_1^2+q_2^2-q_3^2))\big]\times\Cr{13}{\text{SR8}}.
\end{align}
\begin{sloppypar}
The three degrees of freedom can be identified with the three functions $f(q_{1}^{2},q_{2}^{2},q_{3}^{2})$, $g(q_{1}^{2},q_{2}^{2},q_{3}^{2})$, and $h(q_{1}^{2},q_{2}^{2},q_{3}^{2})$, which, apart from constraints imposed by crossing symmetry,  can be chosen freely in order to cancel further singularities in the contribution of tensor states. Their structure can be improved considerably: instead of $\lambda_{123}$ poles, our preferred solution again only exhibits poles in $q_i^2$ (and thus at three single points in the $g-2$ integration region instead of the entire boundary), while for two of the six representative $\hat\Pi_i$ as well as in some special cases in which only some tensor TFFs are non-vanishing all kinematic singularities disappear. We studied systematically whether the remaining three SRs suffice to remove the tensor singularities altogether. However, we find that this is not possible, confirming the statement of Ref.~\cite{Ludtke:2023hvz}. The poles in $q_i^2$ could be traded for singularities of the form $1/(q_i^2-q_j^2)$, corresponding to lines through the $g-2$ integration region, which of course should be avoided, see App.~\ref{app:singularity_structure} for an example. For the same reason, we ignore poles of the type $1/(q_i^2+q_j^2)$, which lie outside that region (for $i\neq j$).
We also checked if kinematic zeros implied by the definite crossing properties of the TFFs could be used to cancel singularities, see Refs.~\cite{Drechsel:1997xv,Colangelo:2015ama,Ludtke:2023hvz} for cases in which such zeros effectively remove Tarrach redundancies, but did not find any further simplifications.
The final basis change is described in more detail in App.~\ref{app:basis_change} and included in the accompanying \textsc{Mathematica} notebook.\footnote{The calculations were performed using \textsc{FeynCalc}~\cite{Mertig:1990an,Shtabovenko:2016sxi,Shtabovenko:2020gxv}.} The $27\times 27$ matrix has determinant $1$, and each non-trivial element is proportional to $s-q_3^2$, since the $g-2$ limit has to remain invariant. 
Importantly, the new $\check\Pi_i$ still fulfill the crossing relations~\eqref{checkPi_crossing}, so that the $u$-channel spectral functions can be inferred from the $s$-channel ones via crossing symmetry as before.
\end{sloppypar}

Our optimized basis enables the evaluation of axial-vector contributions by removing their kinematic singularities and considerably simplifies the representation of tensor contributions. However, as our derivation shows, the complete tensor contributions (and generic contributions of spin $J\ge2$) cannot be fully cured from the problem of kinematic singularities within the dispersive formalism of Ref.~\cite{Colangelo:2017fiz} and therefore require the new dispersive approach of Ref.~\cite{Ludtke:2023hvz}, unless the SRs that cancel the residues are enforced exactly (which is difficult to achieve in practice) or additional SRs are employed (which are difficult to justify in a model-independent way).

\section{Partial-wave convergence of the pion box}
\label{sec:PW}

An important test case for the dispersive formalism is provided by the pion-box contribution, which is defined by two-pion intermediate states whose left-hand cut is again given by pion poles. As shown in Ref.~\cite{Colangelo:2015ama}, this contribution coincides with the scalar-QED loop multiplied by pion form factors for each photon vertex. Since scalar QED fulfills the SRs exactly, so does the pion box, in such a way that a partial-wave expansion has to reproduce the full result for any choice of HLbL basis. To be concrete, we consider as benchmark an evaluation in which the pion form factor is approximated by a $\rho$ pole, $F_\pi^V(q^2)=M_\rho^2/(M_\rho^2-q^2)$, resulting in
\beq
\label{pibox}
a_\mu^{\pi\text{-box}}=-16.42\times 10^{-11}. 
\eeq

\subsection{Sum rules}

\begin{figure}[t]
	\centering
	\includegraphics[width=0.32\linewidth]{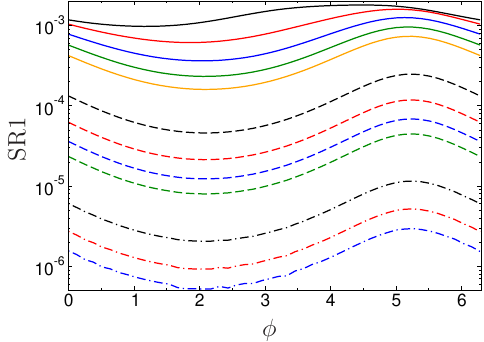}
	\includegraphics[width=0.32\linewidth]{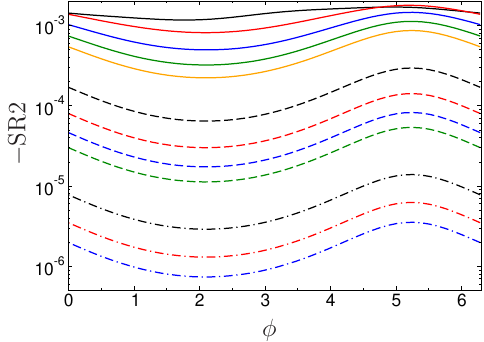}
	\includegraphics[width=0.32\linewidth]{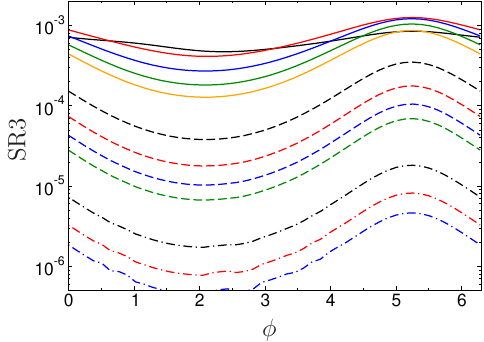}\\
	\includegraphics[width=0.32\linewidth]{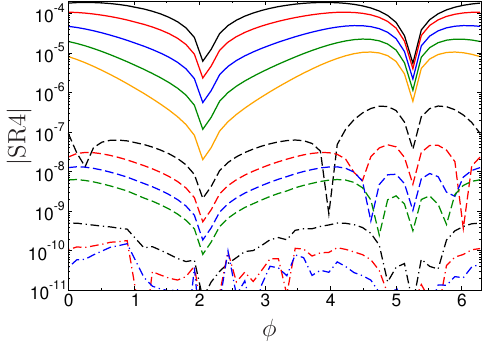}
	\includegraphics[width=0.32\linewidth]{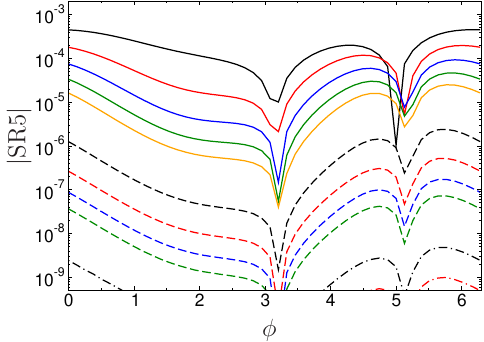}
	\includegraphics[width=0.32\linewidth]{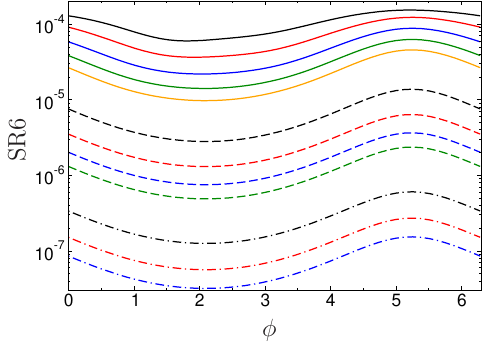}\\
	\includegraphics[width=0.32\linewidth]{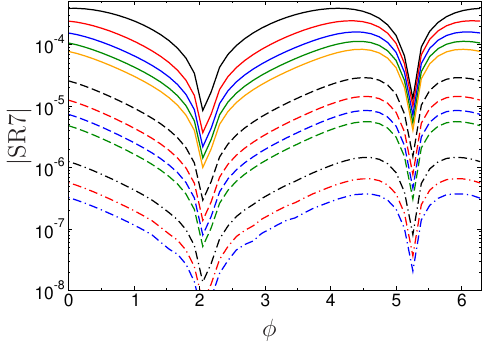}
	\includegraphics[width=0.32\linewidth]{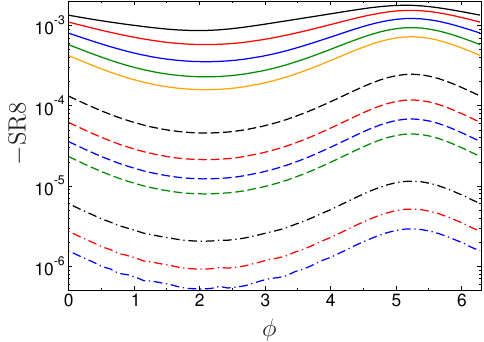}
	\includegraphics[width=0.32\linewidth]{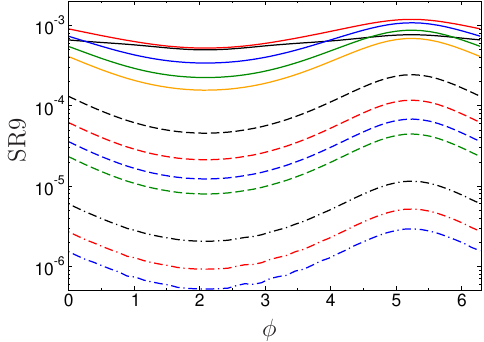}
	\caption{Partial-wave convergence of the nine independent SRs at the point $\Sigma=M_\rho^2$, $r=0.8$, as a function of $\phi$, in the ordering of Eq.~\eqref{eq:SumRulesPicheck}. The solid lines refer to the sum up to $J=2,4,6,8,10$ (black, red, blue, green, orange), the dashed ones up to $J=20,30,40,50$ (black, red, blue, green), and the dot-dashed ones up to $J=100,150,200$ (black, red, blue), respectively. SR4 and SR7 are trivially fulfilled at $\phi\in\big\{\frac{2\pi}{3},\frac{5\pi}{3}\big\}$, where $q_1^2=q_3^2$, since in this case $s$- and $u$-channel contributions cancel.}
	\label{fig:sum_rule_convergence}
\end{figure}

As a first step, we revisited the partial-wave convergence pattern of the SRs. In particular, we implemented partial waves up to $J=200$ to see how fast the convergence occurs over a wide range of parameters. A representative case is shown in Fig.~\ref{fig:sum_rule_convergence}, for all nine independent SRs, at a particular choice of $\Sigma$ and $r$ in the parameterization~\eqref{Qi} of the virtualities, as a function of the angle $\phi$. SR4 and SR7, corresponding to $\check\Pi_{16,24}$, exhibit the trivial zeros at $q_1^2=q_3^2$, otherwise, the rate of convergence does not vary much (with the exception of accidental zeros in SR5), and the same observation pertains to the dependence on $\Sigma$ and $r$. In general, convergence of the SRs seems to proceed roughly logarithmically in $J$, with two orders of magnitude in $J$, from $J=2$ up to $J=200$, suppressing the SR value by about three orders of magnitude, so that reaching high precision does require summing up a large number of partial waves. On the technical level, the evaluation of Legendre functions becomes increasingly unstable for large $J$, we used the implementation from the \textsc{GNU Scientific Library}~\cite{Galassi:2019czg} to produce the results in Fig.~\ref{fig:sum_rule_convergence}.

In practice, we see that the most pronounced cancellations already occur for small values of $J$, most notably between the $++,++$ and $00,++$ helicity amplitudes in the $S$-wave~\cite{Colangelo:2017fiz}, but the convergence pattern observed for the SRs does indicate that saturating the last few percent of the pion-box contribution to $g-2$ may require going to relatively large $J$ already. Accordingly,  it is important to verify that the partial-wave convergence properties for the pion box in our new, optimized basis do not deteriorate compared to Ref.~\cite{Colangelo:2017fiz}.

\subsection[Contribution to $g-2$]{Contribution to $\boldsymbol{g-2}$}
\label{sec:pionbox}

\begin{figure}[t]
	\centering
	\includegraphics[width=0.85\linewidth]{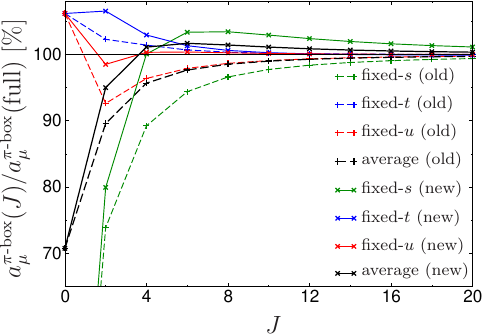}
	\caption{Partial-wave convergence of the pion-box contribution to $a_\mu$ for the basis from Ref.~\cite{Colangelo:2017fiz} (``old,'' dashed lines, plus symbols) and our new basis (solid lines, crosses). In each case, we show fixed-$s$ (green), fixed-$t$ (blue), fixed-$u$ (red), and their average (black). By construction, the two bases yield the same result for $J=0$, at which point the fixed-$s$ contribution vanishes.}
	\label{fig:pion_box_saturation}
\end{figure}

The saturation of the pion-box contribution~\eqref{pibox} for a given angular-momentum cutoff $J$ is shown in Fig.~\ref{fig:pion_box_saturation}, based on the numerical results collected in Table~\ref{tab:pion_box_saturation} (obtained using the \textsc{Cuba} library~\cite{Hahn:2004fe} for the numerical integration).\footnote{Instead of subtracting the integrand at $r=1$ to remove the $1/\lambda_{123}$ pole as in Ref.~\cite{Colangelo:2017fiz}, we separate the integration region into three corridors in $\phi$ and subtract the integrand at $q_i^2=0$, for the virtuality whose pole falls into the respective interval, i.e., for $\phi\in \big[0,\frac{2\pi}{3}\big], \big[\frac{2\pi}{3},\frac{4\pi}{3}\big], \big[\frac{4\pi}{3},2\pi\big]$,  we subtract at $q_1^2,q_3^2,q_2^2=0$. Both subtraction schemes are exact as long as the SRs are fulfilled.} In each case, we include fixed-$(s,t,u)$ representations separately as well as their average, for the basis from Ref.~\cite{Colangelo:2017fiz} and our new basis. In this context, we use ``fixed-$(s,t,u)$'' in the same convention as in Ref.~\cite{Colangelo:2017fiz}, e.g., ``fixed-$s$'' means that a fixed-$s$ dispersion relation is used for each of the representatives $\hat\Pi_{1,4,7,17,39,54}$, while all other required functions are inferred from crossing symmetry. 

By construction, the two bases give identical results for $J=0$, since $S$-waves remain unaffected by the basis change. Therefore, the new basis reproduces the familiar pattern that the fixed-$s$ variant does not contribute at $J=0$, while fixed-$t$ and fixed-$u$ give identical results. Starting from $J=2$ we observe the following differences: the convergence behavior of the fixed-$t$ variant slightly deteriorates, fixed-$u$ improves, and fixed-$s$ first overshoots the full result before converging from the other side. Considering the average, our new basis therefore saturates faster, with $95.0\%$ ($101.1\%$) reached for $J=2$ ($J=4$), compared to $89.6\%$ ($95.7\%$) before. In both cases, the convergence of the last $1\%$ proceeds rather slowly, mainly driven by the fixed-$s$ dispersion relation. We checked that indeed the correct limit is approached, e.g., in our new basis, the fixed-$s$  sum alone has reached $100.3\%$ at $J=40$, implying $100.1\%$ for the average, with a rather slow convergence for the last bit that reflects the convergence pattern observed for the SRs.  

\begin{table}[t]
	\centering
	\renewcommand{\arraystretch}{1.3}
	\begin{tabular}{lcccccccc}
	\toprule
 & \multicolumn{4}{c}{Ref.~\cite{Colangelo:2017fiz} (``old'')} & \multicolumn{4}{c}{This work (``new'')}\\
	$J$ & fixed-$s$ & fixed-$t$ & fixed-$u$ & average & fixed-$s$ & fixed-$t$ & fixed-$u$ & average\\\midrule
 $0$ & $0.00$ & $106.16$ & $106.16$ & $70.78$ & $0.00$ & $106.16$ & $106.16$ & $70.78$ \\
        $2$ & $73.89$ & $102.27$ & $92.61$ & $89.59$ & $79.95$ & $106.52$ & $98.50$ & $94.99$ \\
        $4$ & $89.23$ & $101.40$ & $96.37$ & $95.67$ & $100.06$ & $102.92$ & $100.31$ & $101.09$ \\
        $6$ & $94.38$ & $100.66$ & $97.89$ & $97.64$ & $103.34$ & $101.31$ & $100.36$ & $101.67$ \\
        $8$ & $96.60$ & $100.33$ & $98.65$ & $98.52$ & $103.41$ & $100.60$ & $100.24$ & $101.41$ \\
        $10$ & $97.73$ & $100.17$ & $99.07$ & $98.99$ & $102.90$ & $100.25$ & $100.14$ & $101.10$ \\
        $12$ & $98.37$ & $100.09$ & $99.32$ & $99.26$ & $102.38$ & $100.08$ & $100.08$ & $100.84$ \\
        $14$ & $98.78$ & $100.05$ & $99.49$ & $99.44$ & $101.94$ & $99.98$ & $100.04$ & $100.65$ \\
        $16$ & $99.05$ & $100.03$ & $99.60$ & $99.56$ & $101.59$ & $99.93$ & $100.01$ & $100.51$ \\
        $18$ & $99.24$ & $100.02$ & $99.68$ & $99.65$ & $101.32$ & $99.91$ & $100.00$ & $100.41$ \\ 
        $20$ & $99.38$ & $100.01$ & $99.74$ & $99.71$ & $101.11$ & $99.89$ & $99.99$ & $100.33$\\\bottomrule
	\renewcommand{\arraystretch}{1.0}
	\end{tabular}
	\caption{Partial-wave convergence of the pion-box contribution to $a_\mu$, as illustrated in Fig.~\ref{fig:pion_box_saturation}. The numbers represent the saturation in percent compared to Eq.~\eqref{pibox} when summing up to a given value of $J$.}
	\label{tab:pion_box_saturation}
\end{table}
 
From a practical perspective, it is rather unlikely that any rescattering or resonance contribution with $J>2$ will ever be evaluated, so it is very reassuring that in either basis the error incurred  when truncating at $J=2$, even for the pion box, is very moderate, in our new basis below $1\times 10^{-11}$ and thus phenomenologically irrelevant. In particular, the optimization of the HLbL basis for narrow resonances does not affect the convergence behavior of the pion-box contribution in a negative way, instead, the saturation for low values of $J$ even improves.

\section{Narrow resonances}
\label{sec:narrow_resonances}

By projecting the contribution of the narrow resonances onto the new $\check{\Pi}_i$, it is possible to obtain the corresponding scalar functions $\bar{\Pi}_i$, which form the input for the $g-2$ master formula~\eqref{eq:MasterFormulaPolarCoord}. Using crossing symmetry, only six representatives $\hat{\Pi}_i$ need to be considered, which proceeds as follows: we start from a symmetrized dispersive representation that is the sum of fixed-$(s,t,u)$~\cite{Colangelo:2017fiz}
\begin{align}
\hat{\Pi}_i = \frac{1}{2\pi}&\bigg(\int ds' \frac{\Imst\hat{\Pi}_i(s')}{s'-s}+\int du' \frac{\Imut\hat{\Pi}_i(u')}{u'-u}\notag \\ 
&+\int ds' \frac{\Imsu\hat{\Pi}_i(s')}{s'-s}+\int dt' \frac{\Imtu\hat{\Pi}_i(t')}{t'-t} \notag\\
&+\int dt' \frac{\Imts\hat{\Pi}_i(t')}{t'-t}+\int du' \frac{\Imus\hat{\Pi}_i(u')}{u'-u}\bigg),
\end{align}
where the three lines correspond to fixed-$t$, fixed-$u$, and fixed-$s$, respectively. The factor $1/2$ is needed due to the fact that every pole occurs twice in this symmetrized version. The imaginary part of the scalar functions $\Im\hat{\Pi}_i$ for the different channels can be obtained by exploiting crossing symmetry. Using these crossing properties, we can trace them back to the specific imaginary part $\Imst\hat{\Pi}_i(s')$ of the fixed-$t$ dispersion relation. Exemplarily, we show the respective form for $\hat{\Pi}_1$
\begin{align}
\hat{\Pi}_1 = \frac{1}{2\pi}&\bigg(\int ds' \frac{\Imst\hat{\Pi}_1(s')}{s'-s}+\int du' \frac{\Cr{13}{\Imst\hat{\Pi}_3(u')}}{u'-u} \notag\\
&+\int ds' \frac{\Cr{12}{\Imst\hat{\Pi}_1(s')}}{s'-s}+\int dt' \frac{\Cr{12}{\Cr{13}{\Imst\hat{\Pi}_3(t')}}}{t'-t}\notag \\ 
&+\int dt' \frac{\Cr{23}{\Imst\hat{\Pi}_2(t')}}{t'-t}+\int du' \frac{\Cr{23}{\Cr{13}{\Imst\hat{\Pi}_2(u')}}}{u'-u}\bigg),
\end{align}
where the crossing operator $\mathcal{C}_{ij}$ is defined in Eq.~\eqref{crossing}.
Eventually, in the limit $q_4\rightarrow 0$, we can replace the imaginary parts $\Imst\hat{\Pi}_{i}$ with $\Imst\check{\Pi}_i$.\footnote{To simplify the presentation, we dropped a subtlety in labeling the scalar functions. Strictly speaking, one has for $q_4\to 0$ the identification $\check\Pi_i=\hat\Pi_{g_i}$, $\{ g_i \} = \{1,\ldots, 11, 13, 14, 16, 17, 39, 50, 51, 54\}$, for the $19$ functions relevant for $g-2$, see Ref.~\cite{Colangelo:2017fiz} for more details.} An important prerequisite for this replacement is that the functions $\check{\Pi}_i$, which are only defined for fixed-$t$ kinematics, fulfill the same crossing relations under $\Cr{13}{}$ as the $\hat{\Pi}_i$, see Eq.~\eqref{checkPi_crossing}, which is the main reason to ensure that these crossing relations be maintained in the construction of the new basis. The final form for $\hat{\Pi}_1$ is then given by 
\begin{align}\label{eq:symmetrized_dispersionintegral}
\hat{\Pi}_1\Big|_{q_4\rightarrow 0} = \frac{1}{2\pi}&\bigg(\int ds' \frac{\Imst\check{\Pi}_1(s')}{s'-q_{3}^{2}}+\int du' \frac{\Cr{13}{\Imst\check{\Pi}_3(u')}}{u'-q_{1}^{2}}\notag \\
&+\int ds' \frac{\Cr{12}{\Imst\check{\Pi}_1(s')}}{s'-q_{3}^{2}}+\int dt' \frac{\Cr{12}{\Cr{13}{\Imst\check{\Pi}_3(t')}}}{t'-q_{2}^{2}} \notag\\ 
&+\int dt' \frac{\Cr{23}{\Imst\check{\Pi}_2(t')}}{t'-q_{2}^{2}}+\int du' \frac{\Cr{23}{\Cr{13}{\Imst\check{\Pi}_2(u')}}}{u'-q_{1}^{2}}\bigg).
\end{align}
For narrow resonances, the imaginary parts $\Imst\check{\Pi}_i$ collapse to delta functions $\delta(s-M^2)$, which offset the dispersion integral, leaving a propagator with the respective Mandelstam variable replaced by the corresponding momentum $q_{i}^{2}$. In the following, we will present the results for all six representative scalar functions covering pseudoscalars ($P$), scalars ($S$), axial-vectors ($A$), and tensors ($T$) separately. Analogous results have been presented in Ref.~\cite{Ludtke:2023hvz} for the alternative dispersive approach in triangle kinematics. They differ from our results by terms without resonance pole, reflecting the reshuffling that happens in the transition between the two dispersive approaches~\cite{Ludtke:2023hvz}.

\subsection[Pseudoscalar mesons: $J^{PC}=0^{-+}$]{Pseudoscalar mesons: $\boldsymbol{J^{PC}=0^{-+}}$}

The $T$-matrix element for pseudoscalar mesons is given by
\begin{equation}
\M^{\mu\nu}_P=\epsilon^{\mu\nu\alpha\beta}q_{1\alpha}q_{2\beta}\F_{P\gamma^{\ast}\gamma^{\ast}}(q_{1}^{2},q_{2}^{2}),
\end{equation}
where $\F_{P\gamma^{\ast}\gamma^{\ast}}(q_{1}^{2},q_{2}^{2})$ is the pseudoscalar TFF, fulfilling the crossing relation $\Cr{12}{\F_{P\gamma^{\ast}\gamma^{\ast}}}=\F_{P\gamma^{\ast}\gamma^{\ast}}$. Projecting this onto the imaginary part of the new basis functions $\Imst\check{\Pi}_i(s')$ and using Eq.~\eqref{eq:symmetrized_dispersionintegral}, we obtain 
\begin{align}
\hat{\Pi}_1^P &= \frac{\F_{P\gamma^{\ast}\gamma^{\ast}}(q_{1}^{2},q_{2}^{2})\F_{P\gamma^{\ast}\gamma^{\ast}}(q_{3}^{2},0)}{q_{3}^{2}-\mP^2},\notag\\
\hat{\Pi}_{4}^P&=\hat{\Pi}_{7}^P=\hat{\Pi}_{17}^P=\hat{\Pi}_{39}^P=\hat{\Pi}_{54}^P=0,
\end{align}
while all the remaining ones follow from crossing symmetry. Since pseudoscalar contributions do not depend on the choice for the HLbL basis, this expression is trivially unchanged.

\subsection[Scalar mesons: $J^{PC}=0^{++}$]{Scalar mesons: $\boldsymbol{J^{PC}=0^{++}}$}

The $T$-matrix element for scalar mesons is given by
\begin{equation}
\M^{\mu\nu}_S=\frac{1}{\mS}T_{1}^{\mu\nu}\mathcal{F}_{1}^{S}+\frac{1}{\mS^{3}}T_{2}^{\mu\nu}\mathcal{F}_{2}^{S},
\end{equation}
with the following gauge-invariant structures 
\begin{align}
T_{1}^{\mu\nu}&=q_{1}\cdot q_{2} g^{\mu\nu}-q_{2}^{\nu}q_{1}^{\mu},\notag\\
T_{2}^{\mu\nu}&=q_{1}^{2}q_{2}^{2} g^{\mu\nu}+q_{1}\cdot q_{2} q_{1}^{\mu}q_{2}^{\nu}-q_{1}^{2}q_{2}^{\mu}q_{2}^{\nu}-q_{2}^{2}q_{1}^{\mu}q_{1}^{\nu},
\end{align}
and crossing relations $\Cr{12}{\F_{1,2}^S}=\F_{1,2}^S$.
Following the same procedure as for pseudoscalars, this leads to
\begin{align}
\hat{\Pi}_4^S &= \frac{\big[2\mS^{2}\F_{1}^{S}(q_{1}^{2},q_{2}^{2})-(\mS^{2}+q_{1}^{2}+q_{2}^{2})\F_{2}^{S}(q_{1}^{2},q_{2}^{2})\big]\F_{1}^{S}(q_{3}^{2},0)}{2\mS^{4}(q_{3}^{2}-\mS^{2})},\notag\\
\hat{\Pi}_{17}^S &= \frac{\F_{2}^{S}(q_{1}^{2},q_{2}^{2})\F_{1}^{S}(q_{3}^{2},0)}{\mS^{4}(q_{3}^{2}-\mS^{2})}, \qquad 
\hat{\Pi}_{1}^S=\hat{\Pi}_{7}^S=\hat{\Pi}_{39}^S=\hat{\Pi}_{54}^S=0.
\end{align}
By construction, this result is unchanged from Ref.~\cite{Danilkin:2021icn}.

\subsection[Axial-vector mesons: $J^{PC}=1^{++}$]{Axial-vector mesons: $\boldsymbol{J^{PC}=1^{++}}$}
\label{sec:axialvector}

The $T$-matrix element for axial-vector mesons is given by
\begin{equation}
\M^{\mu\nu\alpha}_A=\frac{i}{\mA^{2}}\sum_{i=1}^{3}T_{i}^{\mu\nu\alpha}\F_{i}^{A}(q_{1}^{2},q_{2}^{2}),
\end{equation}
with the following set of gauge-invariant Lorentz structures 
\begin{equation}
\big\{T_{i}^{\mu\nu\alpha}\big\}=\big\{\epsilon^{\mu\nu\beta\gamma}q_{1 \beta}q_{2 \gamma}(q_{1}^{\alpha}-q_{2}^{\alpha}),\,\epsilon^{\alpha\nu\beta\gamma}q_{1\beta}q_{2\gamma}q_{1}^{\mu}+\epsilon^{\alpha\mu\nu\beta}q_{2\beta}q_{1}^{2},\,\epsilon^{\alpha\mu\beta\gamma}q_{1\beta}q_{2\gamma}q_{2}^{\nu}+\epsilon^{\alpha\mu\nu\beta}q_{1\beta}q_{2}^{2}\big\},
\end{equation}
and crossing relations $\Cr{12}{\F_1^A}=-\F_1^A$, $\Cr{12}{\F_2^A}=-\F_3^A$.
Here we obtain the following results for the six representatives
\begin{align}
\hat{\Pi}_1^A &= \frac{\F_{1}^{A}(q_{3}^{2},0)+\F_{2}^{A}(q_{3}^{2},0)}{2\mA^{6}}\Big[(q_{1}^{2}-q_{2}^{2})\Big(2\F_{1}^{A}(q_{1}^{2},q_{2}^{2})+\F_{a}^{A}(q_{1}^{2},q_{2}^{2})\Big)+(q_{1}^{2}+q_{2}^{2})\F_{s}^{A}(q_{1}^{2},q_{2}^{2})\Big],\notag\\
\hat{\Pi}_4^A &= \frac{\F_{2}^{A}(q_{2}^{2},0)(q_{1}^{2}-q_{2}^{2}+q_{3}^{2})\big[2\F_{1}^{A}(q_{1}^{2},q_{3}^{2})+\F_{3}^{A}(q_{1}^{2},q_{3}^{2})\big]}{2\mA^{4}(q_{2}^{2}-\mA^{2})} + \big(q_1^2\leftrightarrow q_2^2\big),\notag\\ 
 \hat{\Pi}_7^A &= \bigg(\frac{\F_{2}^{A}(q_{2}^{2},0)\big[2\F_{1}^{A}(q_{1}^{2},q_{3}^{2})+\F_{a}^{A}(q_{1}^{2},q_{3}^{2})\big]}{\mA^{4}(\mA^{2}-q_{2}^{2})} 
  - \big(q_1^2\leftrightarrow q_2^2\big)\bigg)-\frac{\F_2^A(q_2^2,0)\F_2^A(q_1^2,q_3^2)}{\mA^4(\mA^2-q_2^2)},\notag\\
\hat{\Pi}_{17}^A &= \frac{\F_{2}^{A}(q_{3}^{2},0)\F_{s}^{A}(q_{1}^{2},q_{2}^{2})}{2\mA^{4}(\mA^{2}-q_{3}^{2})} +\bigg(\frac{\F_{2}^{A}(q_{2}^{2},0)\big[4\F_{1}^{A}(q_{1}^{2},q_{3}^{2})+\F_{a}^{A}(q_{1}^{2},q_{3}^{2})\big]}{2\mA^{4}(\mA^{2}-q_{2}^{2})}+\big(q_1^2\leftrightarrow q_2^2\big)\bigg), \notag\\
\hat{\Pi}_{39}^A &=\sum_{q_i^2\neq q_j^2\neq q_k^2} \hat{\Pi}_{39}^{A, 123}(q_i^2,q_j^2,q_k^2),\qquad \hat{\Pi}_{39}^{A, 123}(q_1^2,q_2^2,q_3^2)=\frac{\F_{2}^{A}(q_{3}^{2},0)\F_{s}^{A}(q_{1}^{2},q_{2}^{2})}{4\mA^{4}(\mA^{2}-q_{3}^{2})},\\ 
\hat{\Pi}_{54}^A &= \frac{\F_{2}^{A}(q_{3}^{2},0)\big[4\F_{1}^{A}(q_{1}^{2},q_{2}^{2})+\F_{a}^{A}(q_{1}^{2},q_{2}^{2})\big]}{2\mA^{4}(q_{3}^{2}-\mA^{2})}+\bigg(\frac{\F_{2}^{A}(q_{2}^{2},0)\F_{s}^{A}(q_{1}^{2},q_{3}^{2})}{2\mA^{4}(q_{2}^{2}-\mA^{2})}-\big(q_1^2\leftrightarrow q_2^2\big)\bigg), \notag
\end{align}
where the (anti-)symmetric TFFs $\F_{s}^{A}$ ($\F_{a}^{A}$) are defined by
\begin{align}
    \F_{s}^{A}(q_{1}^{2},q_{2}^{2}) &= \F_{2}^{A}(q_{1}^{2},q_{2}^{2})-\F_{3}^{A}(q_{1}^{2},q_{2}^{2}),\notag\\
    \F_{a}^{A}(q_{1}^{2},q_{2}^{2}) &= \F_{2}^{A}(q_{1}^{2},q_{2}^{2})+\F_{3}^{A}(q_{1}^{2},q_{2}^{2}).
\end{align}
The result for $\hat{\Pi}_1$ was already anticipated in Ref.~\cite{Colangelo:2021nkr}, the others are new. The expressions have been cast into a form in which their symmetries become manifest, i.e., $\Cr{12}{\hat \Pi_{1,4,17}}=\hat \Pi_{1,4,17}$, $\Cr{12}{\hat \Pi_{54}}=-\hat \Pi_{54}$,  and symmetric under the exchange of either two virtualities for $\hat \Pi_{39}$ (the sum runs over all six permutations).

As previously observed~\cite{Colangelo:2021nkr}, the axial-vector contribution to $\hat\Pi_1$ is a purely polynomial term. In contrast, the axial-vector contributions to the remaining scalar functions contain poles, which in our optimized basis completely agree with the axial-vector contribution defined in triangle kinematics~\cite{Ludtke:2023hvz}, apart from the argument of the singly-on-shell TFF, which here still depends on one photon virtuality instead of being fixed to $M_A^2$.

\subsection[Tensor mesons: $J^{PC}=2^{++}$]{Tensor mesons: $\boldsymbol{J^{PC}=2^{++}}$}
\label{sec:tensor}

The $T$-matrix element for the tensor mesons is given by
\begin{equation}
\M^{\mu\nu\alpha\beta}_T=\sum_{i=1}^{5}T_{i}^{\mu\nu\alpha\beta}\frac{1}{\mT^{n_i}}\F_{i}^{T}(q_{1}^{2},q_{2}^{2}),
\end{equation}
where $n_1 = 1$ and $n_i = 3$ otherwise. The Lorentz structures $T_{i}^{\mu\nu\alpha\beta}$ are defined in Ref.~\cite{Hoferichter:2020lap}, the crossing properties are $\Cr{12}{\F_{1,2,3}^T}=\F_{1,2,3}^T$, $\Cr{12}{\F_{4}^T}=\F_{5}^T$. As discussed in Sec.~\ref{sec:HLbL_tensor}, the freedom in applying SRs is not sufficient to remove all kinematic singularities. Since, moreover, the full expressions become very lengthy, we concentrate here on the important special cases in which the singularities are manifestly absent, and defer the general result to the supplemental  \textsc{Mathematica} notebook. 

First, the projection onto $\hat \Pi_{7,39}$ is free of kinematic singularities for a general choice of tensor TFFs, and the corresponding contributions from $\bar\Pi_{5,6,7,10}$ can therefore be evaluated unambiguously. We find
\begin{align}
	\hat\Pi_7^T &= \bigg[ \F^T(q_1^2) \Big( 2\mT^2 \F_3^T(q_2^2,q_3^2) + (\mT^2-q_2^2-q_3^2) \F_a^T(q_2^2,q_3^2) \Big) - (q_1^2 \leftrightarrow q_2^2) \bigg] \notag\\
		&\quad + \F^T(q_2^2) \bigg[ 2 \Big(\mT^2 \F_1^T(q_1^2,q_3^2) - q_3^2 \F_3^T(q_1^2,q_3^2)\Big) + (\mT^2+q_1^2+3q_3^2) \F_5^T(q_1^2,q_3^2) \bigg] \notag\\
		&\quad + \bigg[ 2 \F^T(q_3^2) (\mT^2-q_1^2 + q_2^2) + \frac{\F_5^T(q_3^2,0)}{\mT^6} \bigg] \Big( \F_2^T(q_1^2,q_2^2) + \F_3^T(q_1^2,q_2^2) - \F_{s}^T(q_1^2,q_2^2)\Big) \notag\\
		&\quad + \frac{\F_5^T(q_1^2,0) \F_3^T(q_2^2,q_3^2)}{\mT^6} - \frac{\F_5^T(q_2^2,0) ( \F_3^T(q_1^2,q_3^2) - \F_5^T(q_1^2,q_3^2) )}{\mT^6} , \notag\\
	\hat \Pi_{39}^T &= \sum_{q_i^2\neq q_j^2\neq q_k^2} \hat{\Pi}_{39}^{T, 123}(q_i^2,q_j^2,q_k^2), \notag\\
	\hat{\Pi}_{39}^{T, 123} &= \Big( \F_4^T(q_1^2,q_2^2) - \F_3^T(q_1^2,q_2^2)\Big)\frac{\F_5^T(q_3^2,0)}{2\mT^6} -
		\bigg[\mT^2 \F_1^T(q_1^2,q_2^2) + (\mT^2-q_1^2) \F_3(q_1^2,q_2^2)\notag\\
		&\quad - \frac{1}{2}\big(3\mT^2-3q_1^2-q_2^2\big)\F_4^T(q_1^2,q_2^2)\bigg]\F^T(q_3^2),
\end{align}
where we defined
\beq
	\F^T(q^2)=\frac{\F_1^T(q^2,0)+\F_5^T(q^2,0)}{\mT^{6}(q^2 - \mT^2)} + \frac{\F_5^T(q^2,0)}{2\mT^8} ,
\eeq
as well as the (anti-)symmetric TFFs $\F_{s}^{T}$ ($\F_{a}^{T}$) 
\begin{align}
    \F_{s}^{T}(q_{1}^{2},q_{2}^{2}) &= \F_{4}^{T}(q_{1}^{2},q_{2}^{2})+\F_{5}^{T}(q_{1}^{2},q_{2}^{2}),\notag\\
    \F_{a}^{T}(q_{1}^{2},q_{2}^{2}) &= \F_{4}^{T}(q_{1}^{2},q_{2}^{2})-\F_{5}^{T}(q_{1}^{2},q_{2}^{2}).
\end{align}

For the other $\hat \Pi_i$ not related to $\hat \Pi_{7,39}$ by crossing symmetry, kinematic singularities are present, in general, but there are two special cases in which they disappear, if either only $\F_{1,3}^T$ or only $\F_{2,3}^T$ are non-vanishing. In these cases we have 
\begin{align}
\hat \Pi_1^T\big|_{\F_{1,3}^T}&=\Big[(q_{1}^{2}-\mT^{2})\F_{3}^{T}(q_{1}^{2},q_{3}^{2})-\mT^{2}\F_{1}^{T}(q_{1}^{2},q_{3}^{2})\Big]\frac{\F_{1}^{T}(q_{2}^{2},0)}{\mT^{6}}+\big(q_1^2\leftrightarrow q_2^2\big),\notag\\
\hat \Pi_4^T\big|_{\F_{1,3}^T}&=\bigg[8\mT^{4}q_{3}^{2}\F_{1}^{T}(q_{1}^{2},q_{2}^{2})+\Big(6\mT^{6}+3\mT^{4}(q_{1}^{2}+q_{2}^{2})-3\mT^{2}(q_{1}^{2}-q_{2}^{2})^{2}\notag \\
&\qquad-q_{3}^{2}\big(4\mT^{4}+5\mT^{2}(q_{1}^{2}+q_{2}^{2})+(q_{1}^{2}-q_{2}^{2})^{2}\big)\Big)\F_{3}^{T}(q_{1}^{2},q_{2}^{2})\bigg]\frac{\F_{1}^{T}(q_{3}^{2},0)}{3\mT^{8}(\mT^{2}-q_{3}^{2})} \notag\\
&+\bigg(\bigg[\Big(\mT^{4}+2\mT^{2}q_{3}^{2}-q_{3}^{2}\big(q_{1}^{2}+2q_{2}^{2}+q_{3}^{2}\big)-q_{1}^{2}q_{2}^{2}\Big)\F_{3}^{T}(q_{1}^{2},q_{3}^{2})\notag\\
&\qquad +\mT^{2}\big(-4\mT^{2}+q_{1}^{2}+3q_{2}^{2}+q_{3}^{2}\big)\F_{1}^{T}(q_{1}^{2},q_{3}^{2})\bigg]\frac{\F_{1}^{T}(q_{2}^{2},0)}{\mT^{6}(\mT^{2}-q_{2}^{2})}+\big(q_1^2\leftrightarrow q_2^2\big)\bigg),\notag\\
\hat \Pi_{17}^T\big|_{\F_{1,3}^T}&=\bigg[\Big(2q_{3}^{2}\big(\mT^{2}+q_{1}^{2}+q_{2}^{2}\big)+3\mT^{2}(q_{1}^{2}+q_{2}^{2})\Big)\F_{3}^{T}(q_{1}^{2},q_{2}^{2}) \notag\\
&\qquad-2\mT^{2}\big(3\mT^{2}+2q_{3}^{2}\big)\F_{1}^{T}(q_{1}^{2},q_{2}^{2})\bigg]\frac{\F_{1}^{T}(q_{3}^{2},0)}{3\mT^{8}(\mT^{2}-q_{3}^{2})}\notag\\
&+\bigg((q_{3}^{2}-q_{1}^{2})\F_{3}^{T}(q_{1}^{2},q_{3}^{2})\frac{\F_{1}^{T}(q_{2}^{2},0)}{\mT^{6}(\mT^{2}-q_{2}^{2})}
+\big(q_1^2\leftrightarrow q_2^2\big)\bigg),\notag\\
\hat \Pi_{54}^T\big|_{\F_{1,3}^T}&=(q_{1}^{2}-q_{2}^{2})\F_{3}^{T}(q_{1}^{2},q_{2}^{2})\frac{\F_{1}^{T}(q_{3}^{2},0)}{\mT^{6}(\mT^{2}-q_{3}^{2})} \notag\\
&+\bigg(\Big[2\mT^{2}\F_{1}^{T}(q_{1}^{2},q_{3}^{2})-(q_{1}^{2}+q_{3}^{2})\F_{3}^{T}(q_{1}^{2},q_{3}^{2})\Big]\frac{\F_{1}^{T}(q_{2}^{2},0)}{\mT^{6}(\mT^{2}-q_{2}^{2})}-\big(q_1^2\leftrightarrow q_2^2\big)\bigg),
\end{align}
and 
\begin{align}
\hat \Pi_1^T\big|_{\F_{2,3}^T}&=q_{1}^{2}\Big(q_{3}^{2}(\mT^{2}+q_{2}^{2})+2q_{2}^{2}(q_{1}^{2}-q_{2}^{2})\Big)\F_{3}^{T}(q_{1}^{2},q_{3}^{2})\frac{\F_{2}^{T}(q_{2}^{2},0)}{2\mT^{8}(q_{1}^{2}+q_{2}^{2})}+\big(q_1^2\leftrightarrow q_2^2\big),\notag\\
\hat \Pi_4^T\big|_{\F_{2,3}^T}&=\bigg[\Big(-\mT^{6}+\mT^{4}\big(q_{1}^{2}+q_{2}^{2}+q_{3}^{2}\big)+\mT^{2}\big(5q_{3}^{2}(q_{1}^{2}+q_{2}^{2})-(q_{1}^{2}-q_{2}^{2})^{2}\big) \notag\\
&\qquad+q_{3}^{2}(q_{1}^{2}-q_{2}^{2})^{2}\Big)\F_{3}^{T}(q_{1}^{2},q_{2}^{2}) \notag\\
&-(\mT^{2}-q_{3}^{2})\Big(\mT^{4}+\mT^{2}(q_{1}^{2}+q_{2}^{2})+(q_{1}^{2}-q_{2}^{2})^{2}\Big)\F_{2}^{T}(q_{1}^{2},q_{2}^{2})\bigg]\frac{\F_{2}^{T}(q_{3}^{2},0)}{6\mT^{10}},\notag\\
\hat \Pi_{17}^T\big|_{\F_{2,3}^T}&=(\mT^{2}-q_{3}^{2})\big(\mT^{2}+q_{1}^{2}+q_{2}^{2}\big)\F_{3}^{T}(q_{1}^{2},q_{2}^{2})\frac{\F_{2}^{T}(q_{3}^{2},0)}{3\mT^{10}}\notag \\
&-\bigg((\mT^{2}+q_{2}^{2})\F_{3}^{T}(q_{1}^{2},q_{3}^{2})\frac{\F_{2}^{T}(q_{2}^{2},0)}{2\mT^{8}}+\big(q_1^2\leftrightarrow q_2^2\big)\bigg),\notag\\
\hat \Pi_{54}^T\big|_{\F_{2,3}^T}&=-(\mT^{2}+q_{2}^{2})\F_{3}^{T}(q_{1}^{2},q_{3}^{2})\frac{\F_{2}^{T}(q_{2}^{2},0)}{2\mT^{8}}-\big(q_1^2\leftrightarrow q_2^2\big),
\end{align}
respectively. In particular, we observe that in the first scenario only terms involving $\F_1^T$ display a pole at $\mT^2$, while the second scenario merely produces polynomial terms. Importantly, these results imply that an evaluation of tensor contributions based on the quark-model TFFs~\cite{Schuler:1997yw}
\beq
\label{quarkmodel}
\frac{\F_1^T(q_1^2,q_2^2)}{\F_1^T(0,0)}\bigg|_{\text{\cite{Schuler:1997yw}}}=\bigg(\frac{\mT^2}{\mT^2-q_1^2-q_2^2}\bigg)^2,\qquad 
 \F_{2,3,4,5}^T(q_1^2,q_2^2)\big|_{\text{\cite{Schuler:1997yw}}}=0,
\eeq
will not be affected by kinematic singularities, which should define a useful benchmark for cases such as the $f_2(1270)$. We remark that the model~\eqref{quarkmodel} reflects the asymptotic scaling of the TFFs derived from the light-cone expansion~\cite{Hoferichter:2020lap} in analogy to Refs.~\cite{Lepage:1979zb,Lepage:1980fj,Brodsky:1981rp}, i.e., while $\F_1^T\propto 1/Q^4$, the other TFFs fall off proportional to $1/Q^6$.

In the framework of Ref.~\cite{Colangelo:2017fiz}, the remaining tensor contributions that contain kinematic singularities cannot be evaluated in a meaningful way (unless one found a way to enforce the SRs exactly by combining the tensors with additional contributions) and instead require the dispersive approach in triangle kinematics~\cite{Ludtke:2023hvz}.  As in the case of axial vectors, we checked again that the residues of the pure tensor-meson poles of our results agree with the tensor-meson contribution defined in Ref.~\cite{Ludtke:2023hvz}, where it was found that all these residues are proportional to $\F_1^T(\mT^2,0) + \F_5^T(\mT^2,0)$.

\section{Summary and outlook}
\label{sec:summary}

In this work we presented a basis for HLbL scattering that is optimized for the evaluation of contributions from narrow resonances. To this end, we used the freedom to perform a basis change as long as the relevant quantities for $g-2$ remain unaltered, as is possible thanks to a set of sum rules fulfilled by the invariant amplitudes of the BTT decomposition. We constructed this new basis to avoid the appearance of kinematic singularities, arising from the transition between BTT and helicity amplitudes, as much as possible: in general, these singularities are only guaranteed to cancel for the entire HLbL tensor, again by virtue of the same set of sum rules, but with a prudent choice of basis their effect can be mitigated.  
Demanding that $S$-waves remain the same, that the singularities be removed from axial-vector contributions, and that the remaining, unavoidable ones for tensor intermediate states become as simple as possible, we find little freedom in the resulting basis. The corresponding basis change~\eqref{basis_change} and the resulting expressions for narrow axial-vector and tensor mesons in Secs.~\ref{sec:axialvector} and~\ref{sec:tensor} constitute our main results. For the latter, we confirm the statement of Ref.~\cite{Ludtke:2023hvz} that in general kinematic singularities cannot be fully removed, leaving poles in the photon virtualities, but we find that important test cases can be evaluated without any ambiguity in subtracting divergences. This includes two of the six representative BTT functions for general TFFs, and scenarios in which only some TFFs are non-vanishing, including the case in which only $\F_1^T$ is present, as predicted by a simple quark model. 

We tested our new basis using the example of the pion-box contribution, to ensure that the optimization for narrow resonances does not impede the partial-wave convergence. Fortunately, we observe little change in the convergence behavior, in fact, the saturation for the lowest partial waves even improves compared to our previous basis. For these reasons, we will adopt the framework established by this new basis as starting point for a future complete dispersive evaluation of the HLbL contribution. Axial-vector resonances can be directly evaluated once input for their TFFs is provided, in such a way that in this case the main challenge becomes assessing potential overlap with and matching to short-distance constraints. For tensor resonances, the new basis allows us to at least provide an estimate for the TFF that from quark-model arguments and its asymptotic scaling would be expected to dominate, defining a valuable benchmark for the complementary dispersive approach in triangle kinematics. Work along all these lines is in progress.

\acknowledgments   
We thank Gilberto Colangelo for useful comments on the manuscript. 
Financial support by the SNSF (Project Nos.\ PCEFP2\_181117 and PCEFP2\_194272) is gratefully acknowledged. 

\appendix

\section{Singularity structure for tensor states}
\label{app:singularity_structure}

Exemplarily, we show the singularity structure of $\Imst\check{\Pi}_3$ for one particular combination of form factors $\F_{1}^{T}(q_{1}^{2},q_{2}^{2})\F_{2}^{T}(q_{3}^{2},0)$
\begin{align}
    \Imst\check{\Pi}_3(s,q_{1}^{2},q_{2}^{2},q_{3}^{2})&=\frac{\left(q_3^2-s\right)\F_{1}^{T}(q_{1}^{2},q_{2}^{2})\F_{2}^{T}(q_{3}^{2},0)}{\mT^{6}q_{1}^{2}}\bigg[ \mT^2 \left(q_2^2-q_3^2\right)+q_1^2 q_3^2 \notag\\
    &\quad +16 q_1^2 q_3^2 \Bigl(\mT^2 \left(q_1^2 \left(q_2^2+q_3^2\right)-\left(q_2^2-q_3^2\right){}^2\right) \notag\\
    &\qquad+q_1^2 q_3^2 \left(-q_1^2-3q_2^2+q_3^2\right) \Bigr) \, g_{3}(q_{1}^{2},q_{2}^{2},q_{3}^{2}) \bigg]\pi\delta(s-\mT^2)+\ldots
\end{align}
Hence out of the three possible degrees of freedom, only one particular function $g_{3}(q_{1}^{2},q_{2}^{2},q_{3}^{2})$ contributes, which ultimately prevents removing the kinematic singularity $1/q_{1}^{2}$. Choosing $g_{3}(q_{1}^{2},q_{2}^{2},q_{3}^{2})=-1/(16q_{1}^{2}(q_{1}^{2}+q_{3}^{2})(q_{3}^{2}-q_{2}^{2}))$ allows one to trade the singularity $1/q_{1}^{2}$ in favor of $1/(q_{3}^{2}-q_{2}^{2})$, but such a singularity would affect lines through the $g-2$ integration region, instead of being concentrated at a single point, and thus lead to a much more complicated subtraction scheme, e.g., for the pion-box contribution.

\section{Basis change}
\label{app:basis_change}
The basis change is described by a $27\times 27$ matrix that has the following form
\beq
    B_{ij}(s,q_{1}^{2},q_{2}^{2},q_{3}^{2})=\delta_{ij}+(s-q_{3}^{2})b_{ij}(q_{1}^{2},q_{2}^{2},q_{3}^{2}).
\eeq
Note that the basis change is derived for fixed-$t$ kinematics at $t=q_{2}^{2}$. Hence the matrix only depends on one Mandelstam variable. In the following, we display the non-zero entries of $b_{ij}(q_{1}^{2},q_{2}^{2},q_{3}^{2})$
\begin{align}
\label{basis_change}
    b_{1,7}&=-\frac{q_{1}^{2}+q_{2}^{2}-q_{3}^{2}}{4q_{3}^{2}}, \quad b_{1,8}=-\frac{q_{1}^{2}+q_{2}^{2}+q_{3}^{2}}{4q_{3}^{2}}, \quad b_{1,11}=b_{1,18}=-b_{1,19}=\frac{q_{1}^{2}-q_{2}^{2}+q_{3}^{2}}{4q_{3}^{2}},\notag \\
    b_{1,12}&=b_{1,13}=-\frac{1}{2},
    \quad b_{1,20}=-2b_{1,24}=\frac{1}{4q_{3}^{2}},\notag \\
    b_{1,22}&=\frac{-q_{1}^{2}+3q_{2}^{2}+q_{3}^{2}}{8q_{3}^{2}\lambda_{123}}, \quad  b_{1,23}=\frac{-q_{1}^{2}+q_{2}^{2}+q_{3}^{2}}{8q_{3}^{2}\lambda_{123}}, \notag\\
    b_{2,7}&=\frac{2q_{1}^{4}q_{2}^{2}-2q_{1}^{2}q_{2}^{4}-8q_{1}^{2}q_{3}^{4}+5q_{1}^{2}q_{2}^{2}q_{3}^{2}+4q_{2}^{2}q_{3}^{4}-q_{2}^{4}q_{3}^{2}}{8q_{2}^{2}q_{3}^{2}(q_{1}^{2}+q_{3}^{2})}, \notag\\
    b_{2,8}&=\frac{2q_{1}^{4}-2q_{1}^{2}q_{2}^{2}+5q_{1}^{2}q_{3}^{2}-q_{2}^{2}q_{3}^{2}}{8q_{3}^{2}(q_{1}^{2}+q_{3}^{2})}, \quad b_{2,9}=-\frac{2q_{1}^{2}+q_{2}^{2}-3q_{3}^{2}}{8(q_{1}^{2}+q_{3}^{2})}, \notag\\ 
    b_{2,10}&=-\Cr{13}{b_{2,9}} ,\quad b_{2,12}=-\Cr{13}{b_{2,7}} ,\quad b_{2,13}=-\Cr{13}{b_{2,8}},\quad b_{2,15}=-\Cr{13}{b_{2,11}},  \notag\\
    b_{2,11}&=-\frac{2q_{1}^{4}-q_{1}^{2}q_{2}^{2}+2q_{1}^{2}q_{3}^{2}+q_{2}^{4}-3q_{2}^{2}q_{3}^{2}}{4q_{2}^{2}(q_{1}^{2}+q_{3}^{2})} ,\quad b_{2,16}=-\frac{q_{1}^{2}-q_{3}^{2}}{8(q_{1}^{2}+q_{3}^{2})}, \notag\\
    b_{2,17}&=-\frac{5q_{1}^{6}+q_{1}^{4}(11q_{3}^{2}-6q_{2}^{2})+q_{1}^{2}\big(q_{2}^{4}-4q_{2}^{2}q_{3}^{2}+11q_{3}^{4}\big)+q_{3}^{2}(q_{2}^{2}-5q_{3}^{2})(q_{2}^{2}-q_{3}^{2})}{8(q_{1}^{2}+q_{3}^{2})\lambda_{123}}, \notag\\
    b_{2,18}&=\frac{1}{8q_2^2 \left(q_1^2+q_3^2\right) \lambda _{123}}\bigg[-4 q_1^8+21 q_2^2 q_1^6+\left(-30 q_2^4+27 q_3^2 q_2^2+8 q_3^4\right) q_1^4\notag\\
    &\qquad+q_2^2 \left(17 q_2^4-36 q_3^2 q_2^2+27 q_3^4\right) q_1^2-\left(q_2^2-q_3^2\right)
   \left(4 q_2^6-13 q_3^2 q_2^4+17 q_3^4 q_2^2-4 q_3^6\right)\bigg], \notag\\
   b_{2,19}&=-\Cr{13}{b_{2,18}}, \quad b_{2,20}=\frac{q_2^2-2 q_1^2}{4 q_2^2 \left(q_1^2+q_3^2\right)}, \quad b_{2,21}=-\Cr{13}{b_{2,20}}, \notag\\
   b_{2,22}&=-\frac{q_1^6\left(q_2^2-6q_3^2\right)+q_1^4q_3^2\left(5q_2^2+4q_3^2\right)-q_1^2\left(q_2^4-q_3^4\right)\left(q_2^2+2q_3^2\right)+q_2^2q_3^2\left(q_2^2-q_3^2\right)^2}{16 q_1^2 q_2^2 q_3^2 \left(q_1^2+q_3^2\right)\lambda _{123}},\notag\\
   b_{2,23}&=-\Cr{13}{b_{2,22}}, \quad b_{2,24}=\frac{\left(q_1^2-q_3^2\right)\left(-q_2^4+q_1^2 q_2^2+q_3^2 q_2^2+2 q_1^2 q_3^2\right)}{16 q_1^2 q_2^2 q_3^2 \left(q_1^2+q_3^2\right)}, \notag\\
   b_{3,7}&=-\Cr{13}{b_{1,12}}, \quad b_{3,8}=-\Cr{13}{b_{1,13}}, \quad b_{3,12}=-\Cr{13}{b_{1,7}}, \quad b_{3,13}=-\Cr{13}{b_{1,8}}, \notag\\
   b_{3,15}&=-\Cr{13}{b_{1,11}}, \quad b_{3,18}=-\Cr{13}{b_{1,19}}, \quad b_{3,19}=-\Cr{13}{b_{1,18}}, \quad b_{3,21}=-\Cr{13}{b_{1,20}}, \notag\\
   b_{3,22}&=-\Cr{13}{b_{1,23}}, \quad b_{3,23}=-\Cr{13}{b_{1,22}}, \quad b_{3,24}=-\Cr{13}{b_{1,24}}, \notag\\
   b_{4,7}&=b_{4,8}=-b_{4,9}=-b_{4,10}=\frac{1}{4}, \quad b_{4,12}=-\frac{2 q_1^2-q_2^2-q_3^2}{4 q_1^2}, \quad b_{4,13}=\frac{2 q_1^2+q_2^2+q_3^2}{4 q_1^2}, \notag\\
   b_{4,15}&=-\frac{3 q_1^2-q_2^2+q_3^2}{4 q_1^2}, \quad b_{4,17}=-\frac{q_2^2 \left(-q_1^2+q_2^2-q_3^2\right)}{\lambda _{123}}, \notag\\
   b_{4,18}&=-b_{4,19}=\frac{3 q_1^6-q_1^4\left(11q_2^2+5q_3^2\right)+q_1^2\left(9q_2^4-10q_2^2q_3^2+q_3^4\right)-\left(q_2^2-q_3^2\right)^3}{4 q_1^2 \lambda _{123}}, \notag\\
   b_{4,21}&=-\frac{1}{4 q_1^2}, \quad b_{4,22}=-\frac{q_1^2+q_2^2-q_3^2}{8 q_1^2 \lambda _{123}}, \quad b_{4,23}=\frac{3 q_1^2-3 q_2^2+q_3^2}{8 q_1^2 \lambda _{123}}, \quad b_{4,24}=\frac{1}{8 q_1^2}, \notag\\
   b_{5,7}&=b_{5,8}=-\frac{2 q_1^4+3 q_3^2 q_1^2+q_3^4-q_2^2 q_3^2}{4 q_3^2 \left(q_1^2+q_3^2\right)}, \quad b_{5,9}=\frac{q_1^2-q_2^2+q_3^2}{4 \left(q_1^2+q_3^2\right)}, \notag\\
   b_{5,10}&=-\Cr{13}{b_{5,9}}, \quad b_{5,12}=b_{5,13}=-\Cr{13}{b_{5,7}}, \notag\\
   b_{5,17}&=-\frac{\left(q_1^2+q_2^2+q_3^2\right) \left(-q_2^4+q_1^2 q_2^2+q_3^2 q_2^2+8 q_1^2 q_3^2\right)}{4\left(q_1^2+q_3^2\right) \lambda _{123}}, \notag\\
   b_{5,18}&=-b_{5,19}=\frac{\left(q_1^2+q_2^2+q_3^2\right) \left(-q_2^4+q_1^2 q_2^2+q_3^2 q_2^2+8 q_1^2 q_3^2\right)}{4\left(q_1^2+q_3^2\right) \lambda _{123}}, \notag\\
   b_{5,22}&=\frac{\left(q_1^2-q_3^2\right) \left(q_1^2+q_2^2-q_3^2\right)}{8 q_1^2 q_3^2 \lambda _{123}}, \quad b_{5,23}=-\Cr{13}{b_{5,22}}, \quad b_{5,24}=-\frac{q_1^2-q_3^2}{8 q_1^2 q_3^2}, \notag\\
   b_{6,7}&=-\Cr{13}{b_{4,12}}, \quad b_{6,8}=-\Cr{13}{b_{4,13}}, \quad b_{6,9}=-\Cr{13}{b_{4,10}} , \quad b_{6,10}=-\Cr{13}{b_{4,9}} ,\notag\\
   b_{6,11}&=-\Cr{13}{b_{4,15}} , \quad b_{6,12}=-\Cr{13}{b_{4,7}}, \quad b_{6,13}=-\Cr{13}{b_{4,8}}, \notag\\
   b_{6,17}&=\Cr{13}{b_{4,17}}, \quad b_{6,18}=-\Cr{13}{b_{4,19}}, \quad b_{6,19}=-\Cr{13}{b_{4,18}}, \quad b_{6,20}=-\Cr{13}{b_{4,21}}, \notag\\
   b_{6,22}&=-\Cr{13}{b_{4,23}}, \quad b_{6,23}=-\Cr{13}{b_{4,22}}, \quad b_{6,24}=-\Cr{13}{b_{4,24}}, \notag\\
   b_{11,7}&=b_{11,8}=\frac{1}{2 q_3^2}, \quad b_{11,12}=b_{11,13}=\frac{1}{2 q_1^2}, \notag\\
   b_{11,22}&=-\frac{q_1^2+q_2^2-q_3^2}{8 q_1^2 q_3^2 \lambda _{123}}, \quad b_{11,23}=\frac{q_1^2-q_2^2-q_3^2}{8 q_1^2 q_3^2 \lambda _{123}}, \quad b_{11,24}=\frac{1}{8 q_1^2 q_3^2}, \notag\\
   b_{14,7}&=\frac{q_2^2-q_3^2}{2 q_2^2 q_3^2}, \quad b_{14,8}=\frac{1}{2 q_3^2}, \quad b_{14,12}=-\Cr{13}{b_{14,7}}, \quad b_{14,13}=-\Cr{13}{b_{14,8}}, \notag\\
   b_{14,11}&=\frac{-q_1^2+q_2^2-q_3^2}{4 q_2^2 q_3^2}, \quad b_{14,15}=-\Cr{13}{b_{14,11}}, \notag\\
   b_{14,18}&=-\frac{\left(q_1^2+q_3^2\right) \left(q_1^2-q_2^2+q_3^2\right)}{4 q_1^2 q_2^2 q_3^2}, \quad b_{14,19}=-\Cr{13}{b_{14,18}}, \notag\\
   b_{14,20}&=-\frac{1}{4 q_2^2 q_3^2}, \quad b_{14,21}=-\Cr{13}{b_{14,20}}, \notag\\
   b_{14,22}&=\frac{3q_1^6-q_1^4\left(6q_2^2-q_3^2\right)-q_1^2\left(q_2^4+2q_2^2q_3^2+3q_3^4\right)+q_3^2\left(q_2^4-q_3^4\right)}{16 q_1^2 q_2^2 q_3^2 \left(q_1^2+q_3^2\right)\lambda _{123}}, 
   \quad b_{14,23}=-\Cr{13}{b_{14,22}}, \notag\\
   b_{14,24}&=\frac{\left(q_1^2-q_3^2\right) \left(q_1^2+q_2^2+q_3^2\right)}{16 q_1^2 q_2^2 q_3^2 \left(q_1^2+q_3^2\right)}, \notag\\
   b_{15,7}&=-\Cr{13}{b_{11,12}}, \quad b_{15,8}=-\Cr{13}{b_{11,13}}, \quad b_{15,12}=-\Cr{13}{b_{11,7}}, \quad b_{15,13}=-\Cr{13}{b_{11,8}}, \notag\\
   b_{15,22}&=-\Cr{13}{b_{11,23}}, \quad b_{15,23}=-\Cr{13}{b_{11,22}}, \quad b_{15,24}=-\Cr{13}{b_{11,24}}, \notag\\
   b_{17,7}&=b_{17,8}=-\frac{1}{2 q_3^2}, \quad b_{17,12}=b_{17,13}=\Cr{13}{b_{17,7}}, \notag\\
   b_{17,22}&=\frac{q_1^2+q_2^2-q_3^2}{8 q_1^2 q_3^2 \lambda _{123}}, \quad b_{17,23}=\Cr{13}{b_{17,22}}, \quad b_{17,24}=-\frac{1}{8 q_1^2 q_3^2}, \notag\\
   b_{18,7}&=-\frac{q_2^2-q_3^2}{2 q_2^2 q_3^2}, \quad b_{18,8}=-\frac{1}{2 q_3^2}, \quad b_{18,12}=-\frac{1}{2 q_2^2}, \notag\\
   b_{18,11}&=-\frac{-q_1^2+q_2^2-q_3^2}{4 q_2^2 q_3^2}, \quad b_{18,15}=-\frac{q_1^2-q_2^2+q_3^2}{4 q_1^2 q_2^2}, \notag\\
   b_{18,18}&=-b_{18,19}=\frac{\left(q_1^2+q_3^2\right) \left(q_1^2-q_2^2+q_3^2\right)}{4 q_1^2 q_2^2 q_3^2}, \quad b_{18,20}=\frac{1}{4 q_2^2 q_3^2}, \quad b_{18,21}=-\frac{1}{4 q_1^2 q_2^2}, \notag\\
   b_{18,22}&=-\frac{3 q_1^4-2q_1^2\left(3q_2^2+q_3^2\right)-\left(q_2^2-q_3^2\right)^2}{16 q_1^2 q_2^2 q_3^2 \lambda _{123}}, 
   \quad b_{18,23}=-\frac{\left(q_1^2+q_2^2-q_3^2\right) \left(q_1^2-q_2^2+3 q_3^2\right)}{16 q_1^2 q_2^2 q_3^2 \lambda _{123}}, \notag\\
   b_{18,24}&=-\frac{q_1^2+q_2^2-q_3^2}{16 q_1^2 q_2^2 q_3^2}, \notag\\
   b_{19,7}&=-\Cr{13}{b_{18,12}}, \quad b_{19,12}=-\Cr{13}{b_{18,7}}, \quad b_{19,13}=-\Cr{13}{b_{18,8}}, \notag \\
   b_{19,11}&=-\Cr{13}{b_{18,15}}, \quad b_{19,15}=-\Cr{13}{b_{18,11}}, \quad b_{19,18}=-\Cr{13}{b_{18,19}},\notag\\
   b_{19,19}&=-\Cr{13}{b_{18,18}}, \quad
   b_{19,20}=-\Cr{13}{b_{18,21}}, \quad b_{19,21}=-\Cr{13}{b_{18,20}}, \notag\\
   b_{19,22}&=-\Cr{13}{b_{18,23}}, \quad b_{19,23}=-\Cr{13}{b_{18,22}}, \quad
   b_{19,24}=-\Cr{13}{b_{18,24}}.
\end{align}

\bibliographystyle{apsrev4-1_mod_2}
\bibliography{ref}
	
\end{document}